\documentclass{aa}

\usepackage{natbib}
\bibpunct{(}{)}{;}{a}{}{,}

\usepackage{graphicx}
\usepackage{txfonts}
\usepackage[colorlinks=true, citecolor=blue, linkcolor=blue, urlcolor=blue]{hyperref}

\usepackage{color}
\usepackage{threeparttable}
\usepackage{multirow}

\begin{document}
    \title{Rate of repeating tidal disruption events with 5-19 years interval}
   \author{Yujun Yao\inst{1}, 
           Luming Sun\inst{1},
           Tao Wu\inst{1},
           Shiyan Zhong\inst{2},
           Ning Jiang\inst{3,4},
           and Xinwen Shu\inst{1}
           }
   \authorrunning{Y.-J. Yao et al.}
   \institute{Department of Physics, Anhui Normal University, Wuhu, Anhui 241002, China. \\
             \email{sunluming@ahnu.edu.cn}
         \and
             South-Western Institute for Astronomy Research, Yunnan University, Kunming, 650500 Yunnan, China
         \and
             CAS Key Laboratory for Researches in Galaxies and Cosmology, Department of Astronomy, University of Science and Technology of China, Hefei, Anhui 230026, China
         \and
             School of Astronomy and Space Sciences, University of Science and Technology of China, Hefei, Anhui 230026, China
             }

   \date{Received 26 January 2026 / Accepted 25 March 2026}

   \abstract
   {Statistics on tidal disruption events (TDEs) may be contaminated by repeating TDEs (rTDEs), which have been widely discovered recently.
   However, no statistical study on rTDEs with time intervals $>5$ years has been made yet.
   In addition, the origin of rTDEs remains unclear.}
   {We aim to search for rTDEs with time intervals of $>5$ years in a well-defined TDE sample, and estimate the rTDE rate and fraction in the sample.}
   {Using a sample of 16 TDEs at $z<0.05$ from the Zwicky Transient Facility (ZTF) Bright Transient Survey (BTS), we searched for flares 5--19 years before the ZTF TDEs using the Catalina Real-time Sky Survey (CRTS) light curves.
   To distinguish between TDEs and supernovae (SNe), we analyzed archival multi-band data and estimated the expected number of SNe that CRTS could detect in the sample.}
   {We found 2 rTDE candidates, AT 2019azh and AT 2024pvu, with time intervals of 13.2 and 17.1 years, respectively.
   The peak luminosities of CRTS flares are close to those of ZTF flares.
   For the CRTS flare of AT 2024pvu, using the Galaxy Evolution Explorer (GALEX) UV observations near the peak, we measured a blackbody temperature of $\sim19500$ K, consistent with TDEs and higher than SNe.
   Moreover, we estimated the expected number of SNe in the sample to be $\lesssim0.08$, and hence the probability that both CRTS flares are SNe is only 0.3\%.
   Therefore, the possibility that both CRTS flares are SNe can be ruled out, and it is likely that both are TDEs.
   Using the two rTDEs, we inferred that the TDE rate is 2--3 orders of magnitude higher than the average over 5--19 years prior to TDE detection.
   Considering another two rTDEs with intervals of $\sim$2 years in the sample and possible rTDEs missed by CRTS, rTDEs with intervals of $<20$ years may account for 25\%--60\% of the TDE sample.
   We prefer to explain rTDEs as repeating partial TDEs.
   If so, the high fraction of rTDEs suggests that the observed optical TDE rate has been overestimated.
   However, the possibility of independent TDEs cannot be ruled out and requires future observational tests.
   }
   {}
   \keywords{galaxies: nuclei -- methods: statistical}

   \maketitle
   \nolinenumbers

\section{Introduction}

A tidal disruption event (TDE) occurs when a star approaches the tidal radius ($R_t$) of a supermassive black hole (SMBH) and is tidally disrupted \citep[e.g.,][]{Hills1975,Rees1988}:
\begin{equation}
    R_t \approx R_\star \left( \frac{M_{\rm BH}}{M_\star} \right)^{1/3}
\end{equation}
where $R_\star$ and $M_\star$ are the radius and mass of the star, and $M_{\rm BH}$ is the mass of the SMBH.
TDE can be observed if it occurs outside the event horizon, which sets an upper limit for $M_{\rm BH}$ of $\sim 3 \times 10^8 M_{\odot}$ for Schwarzschild SMBHs and Sun-like stars.
Below this limit, TDE produces a bright UV-optical and/or X-ray flare \citep{Komossa2015,van2020,Saxton2020} with a light curve roughly following $L\propto t^{-5/3}$ in theory \citep{Rees1988}.
Its optical spectrum is characterized by a blue continuum with a high and nearly constant blackbody temperature \citep[e.g.,][]{Arcavi2014,Gezari2021,Yao2023}, different from other impostors such as supernovae (SNe) or active galactic nuclei (AGN) flares \citep{Zabludoff2021}.

TDE provides a unique opportunity to discover and statistically analyze inactive SMBHs \citep[e.g.,][]{Stone2016,van2018}, especially at the low-mass end where the classical dynamical method encounters difficulties \citep{Mockler2026}.
Thanks to large-area high-cadence optical surveys such as the All Sky Automated Survey for Supernovae \citep[ASAS-SN,][]{Shappee2014,Kochanek2017}, the Asteroid Terrestrial impact Last Alert System \citep[ATLAS,][]{Tonry2018,Smith2020,Shingles2021} and the Zwicky Transient Facility \citep[ZTF,][]{Bellm2019,Masci2019}, and X-ray surveys such as eROSITA \citep{Predehl2021}, the number of TDEs discovered has been growing rapidly in recent years to $\gtrsim10^2$ \citep[e.g.,][]{van2021,Hammerstein2023,Sazonov2021,Zhang2025}.
The recent rate estimates of optical and X-ray TDEs are $3.1^{+0.6}_{-1.0}$ and $2.3^{+1.2}_{-0.9}\ \times10^{-7}$ $\rm Mpc^{-3}\ yr^{-1}$, respectively, or $\sim2-3 \times 10^{-5}\ \rm galaxy^{-1}\ yr^{-1}$ \citep[e.g.,][]{Yao2023,Grotova2025}.
The total TDE rate, after taking into account dust obscured TDEs revealed in the mid-infrared (MIR) bands \citep[$\sim2\times10^{-5}\ \rm galaxy^{-1}\ yr^{-1}$,][]{Jiang2021_rate,Masterson2024,Yao2025_rate}, roughly matches the prediction of $10^{-5}-10^{-4}\ \rm galaxy^{-1}\ yr^{-1}$ \citep[e.g.,][]{Wang2004,Stone2020} by the loss cone theory \citep[e.g.,][]{Frank1976,Merritt2013}.

The above rate statistics are the average values among all galaxies.
However, the TDE rate in galaxies of different types and properties can vary greatly \citep[e.g.,][]{Chang2025,Hannah2025}.
Observationally, TDEs are preferably hosted in post-starburst galaxies with a rate magnified by a factor of several to several tens \citep[e.g.,][]{Liu2013_TDErate,Arcavi2014,French2016,Smith2017,Yao2023}.
The origin of the preference remains unclear \citep[see details in][]{Stone2020}.

Another possible problem with TDE rate statistics arises from repeating TDEs (rTDEs), which have been widely observed in recent years.
The first candidate was found in the Seyfert galaxy IC 3599, with one flare detected in X-ray in 1990 \citep{IC3599_B1995,IC3599_G1995}, and another flare in both the optical and X-ray bands in 2009 \citep{IC3599_C2015,IC3599_G2015}, although the TDE nature is under debate \citep{IC3599_S2015}.
The first periodic rTDE is ASASSN-14ko, which has $>30$ flares to date with a period of 114 days \citep{Payne2021,Payne2022,Payne2023,Huang2025}.
Other rTDEs or candidates include AT 2018fyk \citep{Wevers2023}, AT 2020vdq \citep{Somalwar2025}, AT 2022dbl \citep{Lin2024,Hinkle2024_2022dbl}, eRASSt J045650.3‒203750 \citep{Liu2023,Liu2024}, IRAS F01004-2237 \citep{Sun2024}, AT 2021aeuk \citep{Bao2024,Sun2025}, AT 2022sxl \citep{Ji2025} and AT 2023uqm \citep{Wang2025}.

The origin of rTDE is still unclear.
A widely accepted model, repeating partial TDE \citep[rpTDE,][]{Payne2021}, assumes that a single star in an elliptical orbit around an SMBH undergoes partial disruption repeatedly.
The most substantial evidence comes from the good periodicity of flares in ASASSN-14ko and AT 2023uqm (period of 527 days).
The semi-major axis $a$ of the star derived from the period is $\sim2$ orders of magnitude larger than $R_t$, requiring a highly eccentric orbit.
Stars with such orbits can be produced through the Hills mechanism \citep[e.g.,][]{Hills1988}, although the details are still unclear.
To explain the generation of multiple flares, the model requires that the star be partially disrupted with its core remains \citep[e.g.,][]{Ryu2020_I,Ryu2020_III}.
This in turn demands that the star's penetration factor ($\beta=R_t/R_p$, where $R_p$ is the periapsis distance) is $<1$, unlike the classic fully disrupted events with $\beta\geq1$.
Besides the rpTDE model, there are other models such as independent TDEs \citep{Sun2024}, double TDEs \citep{Mandel2015} and models involving SMBH binary \citep[SMBHB,][]{Coughlin2018}.

Statistical analysis of rTDEs may be essential for studying the TDE rate.
On one hand, if it can be verified that different TDEs observed in one galaxy are independent, the TDE rate therein will be extremely high ($\sim10^{-2}-10^{-1}$ $\rm galaxy^{-1}\ yr^{-1}$).
This will confirm predictions from some theories, for example that a close SMBH companion can boost the TDE rate by several orders of magnitude \citep{Ivanov2005}.
On the other hand, if multiple TDE flares are physically related, the TDE rate is overestimated because these flares should be regarded as one event in statistics.
If rTDEs account for a significant fraction of current TDE samples, they will seriously contaminate research on the TDE rate and preferences.

\citet{Somalwar2025} explored the fraction of rTDE for the first time based on a complete ZTF sample of \citet{Yao2023} containing 33 TDEs.
They found one rTDE using the ZTF data and inferred a fraction of 1/33 with a $3\sigma$ upper limit of $\lesssim40\%$.
However, this fraction may be underestimated because they did not account for rTDEs with periods of $>5$ years.

The rTDEs with longer time intervals can be studied using historical data such as the Catalina Real-time Sky Survey \citep[CRTS,][]{CRTS2009}, which began in 2005 and allows the study of rTDEs with intervals up to $\sim20$ years.
Using CRTS data, \citet{Hinkle2021} and \citet{Langis2025} noticed two rTDE candidates with time intervals of $>10$ years, AT 2019azh and AT 2024pvu.
However, there are no statistical analysis studies on long-interval rTDEs yet.
Furthermore, whether these two events are rTDEs remains unclear, and the possibility of a combination of TDE and SN, as in AT 2021mhg \citep[TDE+SN Ia,][]{Somalwar2025}, cannot yet be ruled out.

In this work, we conducted a systematic search for rTDE candidates using CRTS data for a well-defined TDE sample, verified their nature, and estimated the fraction and rate of rTDEs.
The paper is structured as follows.
In section 2, we describe how we selected rTDE candidates and probe the nature of the CRTS flare of AT 2024pvu using multi-band data.
In section 3, we discuss the possibility of SN contamination, estimate the fraction and rate of rTDE, and explore the possible origin.
We finally summarize the results in Section 4.
Throughout this paper, we assumed the cosmological constants of $H_0 = 70\ {\rm km\ s^{-1}\ Mpc^{-1}}$, $\Omega_M = 0.3$, $\Omega_\Lambda = 0.7$, $T_{\rm cmb} = 2.725\ {\rm K}$, and calculated the luminosity distance using \texttt{cosmology} in \texttt{Astropy} \citep{astropy} based on redshift.

\defcitealias{van2021}{van21}
\defcitealias{Yao2023}{Yao23}

\section{Sample selection and data analysis}
\subsection{Selection of rTDE candidates} \label{sec:2.1}

\begin{table*}
\scriptsize
\centering
\caption{Summary of 16 ZTF BTS TDEs at $z<0.05$}
\begin{tabular}{cccccccccccc}
\hline
\hline
IAU ID     & RA          & Dec         & $z$      & t$_{\rm peak}$& ref &Time interval & $M _{V,\rm limit}$& lg$M_\star$            & lg$\dot{M}_\star$      & EMD$_{\rm SN Ia}$& EMD$_{\rm TDE}$\\
(1)        & (2)         & (3)         & (4)      & (5)           & (6) & (7)          & (8)               & (9)                    & (10)                   & (11)             &(12)            \\ \hline
AT 2018hyz & 10:06:50.87 & +01:41:34.0 & 0.04573  & 58425         & 1   & 5.2-13.0     & -17.91            & 9.99 $^{+0.13}_{-0.18}$& \textless{}-1.84       & 0.02-2.77        & 3.31-5.44      \\
AT 2019azh & 08:13:16.95 & +22:38:53.9 & 0.022    & 58561         & 2   & 5.1-13.6     & -18.34            & 10.30$^{+0.11}_{-0.15}$& \textless{}-1.58       & 0.00-4.34        & 4.34-6.23      \\
AT 2019qiz & 04:46:37.88 & -10:13:34.8 & 0.0151   & 58767         & 2   & 5.9-13.8     & -17.77            & 10.27$^{+0.09}_{-0.11}$& \textless{}-0.93       & 0.10-2.53        & 3.08-5.15      \\
AT 2020vdq & 10:08:53.45 & +42:43:00.3 & 0.044    & 59113         & 2,3 & 7.0-14.6     & -17.05            & 9.32 $^{+0.13}_{-0.18}$& \textless{}-0.77       & 0.81-2.49        & 4.26-7.37      \\
AT 2020wey & 09:05:25.88 & +61:48:09.2 & 0.02738  & 59156         & 2   & 6.8-14.4     & -16.71            & 9.79 $^{+0.09}_{-0.12}$& \textless{}-1.52       & 0.86-2.05        & 3.69-6.73      \\
AT 2020vwl & 15:30:37.81 & +26:58:56.7 & 0.0325   & 59167         & 2   & 7.2-15.1     & -17.46            & 10.02$^{+0.08}_{-0.10}$& \textless{}-1.5        & 0.91-2.76        & 3.63-6.47      \\
AT 2021ehb & 03:07:47.81 & +40:18:40.8 & 0.018    & 59315         & 2   & 7.3-15.2     & -17.88            & 10.32$^{+0.07}_{-0.08}$& \textless{}-1.62       & 0.00-0.50        & 0.77-1.86      \\
AT 2021nwa & 15:53:51.28 & +55:35:19.6 & 0.047    & 59403         & 2   & 7.4-14.5     & -17.90            & 10.23$^{+0.10}_{-0.12}$& \textless{}-1.33       & 0.01-0.91        & 1.20-2.74      \\
AT 2022bdw & 08:25:10.35 & +18:34:57.5 & 0.03782  & 59628$^*$     & -   & 7.9-16.3     & -18.83            & 10.49$^{+0.10}_{-0.14}$& 0.17$^{+0.05}_{-0.06}$ & 0.00-3.62        & 2.15-4.61      \\
AT 2022dbl & 12:20:45.07 & +49:33:04.6 & 0.0284   & 59638         & 4   & 8.5-16.3     & -17.77            & 10.23$^{+0.06}_{-0.07}$& \textless{}-3.1        & 0.10-1.46        & 1.97-3.88      \\
AT 2022lri & 02:20:08.01 & -22:43:15.2 & 0.03275  & 59682         & 5   & 8.2-16.2     & -17.72            & 9.78 $^{+0.12}_{-0.16}$& \textless{}-1.99       & 0.37-1.83        & 2.62-5.27      \\
AT 2022gri & 07:18:20.77 & +33:59:41.5 & 0.028    & 59820$^*$     & -   & 8.7-16.9     & -18.00            & 10.48$^{+0.08}_{-0.10}$& \textless{}-2.04       & 0.01-3.14        & 3.70-6.06      \\
AT 2018meh & 11:40:09.40 & +15:19:38.5 & 0.011    & 59977         & 6,7 & 9.5-17.6     & -17.46            & 10.14$^{+0.08}_{-0.09}$& -0.26$^{+0.06}_{-0.06}$& 1.19-3.77        & 4.78-7.54      \\
AT 2023mhs & 13:43:15.66 & +19:15:00.9 & 0.0482   & 60135$^*$     & -   & 9.6-17.4     & -18.33            & 10.25$^{+0.13}_{-0.19}$& \textless{}-0.27       & 0.00-0.72        & 0.66-2.70      \\
AT 2020afhd& 03:13:35.68 & -02:09:06.2 & 0.027    & 60353$^*$     & -   & 10.0-17.9    & -17.80            & 10.23$^{+0.10}_{-0.13}$& \textless{}-0.74       & 0.13-2.55        & 3.13-5.21      \\
AT 2024pvu & 23:31:11.92 & +22:15:31.9 & 0.048    & 60546$^*$     & -   & 10.3-18.3    & -18.89            & 10.51$^{+0.10}_{-0.14}$& \textless{}-0.72       & 0.00-0.95        & 0.23-1.33      \\
\hline
\hline
\end{tabular}
\begin{tablenotes}
    \item (1)-(4): The IAU IDs, coordinates and redshifts of the ZTF TDEs from the BTS catalog.
    Not that AT 2018meh has a more common name AT 2023clx.
    (5),(6): The peak time and the reference.
    The references are: 1. \citealt{Hammerstein2023}; 2. \citealt{Yao2023}; 3. \citealt{Somalwar2025}; 4. \citealt{Lin2024}; 5. \citealt{2022lri}; 6. \citealt{2023clx_1}; 7. \citealt{2023clx_2}.
    The $*$ label represents t$_{\rm peak}$ taken from the BTS catalog for TDEs with no available peak time from the literature.
    For AT 2020vdq and AT 2022dbl with two ZTF flares, we adopted the time of the earlier peak.
    For AT 2022gri, the $t_{\rm peak}$ in BTS is inaccurate; thus, we re-estimated it visually from the light curve, with a precision of $\sim10$ days.
    (7): The time differences between the CRTS observations and the ZTF TDEs' peak time in units of years.
    (8): The median $5\sigma$ limit magnitude (converted to $M_V$) of the CRTS light curve.
    (9),(10): The stellar mass and star formation rate of the host galaxy from CIAGLE, in units of $M_\odot$ and $M_\odot\ {\rm yr}^{-1}$, respectively.
    (11),(12): The effective monitoring durations in unit of year, assuming SN Ia-like light curves with peak $M_V$ ranging from $-18$ to $-19.4$, and assuming TDE-like light curves with peak $L_V$ of $10^{43} {\rm erg}$ ${\rm s^{-1}}$ with different rise and decline time scales.
\end{tablenotes}
\label{tab1}
\end{table*}

We started from the TDE catalog\footnote{We selected 71 TDEs before July 1, 2025; \url{https://sites.astro.caltech.edu/ztf/bts/bts.php} } of the ZTF Bright Transient Survey \citep[BTS,][]{BTS2020_1,BTS2020_2,BTS2024_3}.
We adopted a redshift cut of $z<0.05$ to increase the completeness of detection for TDE flares by shallower CRTS, and selected 18 ZTF TDEs.
One TDE in the sample, AT 2022wtn, was hosted by a galaxy merger, whose center position in the CRTS catalog was 7$\arcsec$ away from the TDE position.
In this case, any potential flare at the TDE position would be unlikely to be detected in the CRTS light curve, so we removed AT 2022wtn from the sample.
We also removed AT 2024tvd, which was reported to be off-nuclear \citep{2024tvd_1}.
The final sample contains 16 TDEs, whose information is listed in Table~\ref{tab1}.
We also list the time differences between the CRTS observations and the ZTF TDE's peak time, as well as the depth of CRTS (expressed as the $V$-band absolute magnitude $M_{\rm V,limit}$ for $5\sigma$ detection).
CRTS data can detect rTDEs with time intervals of 5 to 19 years with depths ($M_{\rm V,limit}$) ranging from $-16.7$ to $-18.9$ in the sample.

We obtained the CRTS single-exposure photometry of their host galaxies from the CRTS data release 2\footnote{\url{http://nesssi.cacr.caltech.edu/DataRelease/}}, which includes data taken between 2005 and 2013.
CRTS typically took a sequence of 4 consecutive exposures for a source during one night.
The photometry from these exposures should show no significant differences, so we binned them into a single data point.
Before binning, we checked for any abnormal photometric data by examining whether magnitudes were consistent within the margin of error: we required that the standard deviation of the magnitudes be less than 3 times the root-mean-square of the magnitude errors; otherwise, we discarded all photometry for this night.
We adopted the weighted average of the magnitudes as the final magnitude of this night, where the weight was the reciprocal of the square of the magnitude error, and the final magnitude error was calculated using the rule of error propagation.

CRTS took images without a filter and calibrated the photometric data to a pseudo-V magnitude ($V_{\rm CSS}$) close to Bessel/$V$ magnitude.
The difference is related to color and is expressed as \citep{Drake2013}:
\begin{equation}
V = V_{\rm CSS} + 0.31 \times (B-V)^2 + 0.04
\end{equation}
In this work, we studied TDEs with blue $B-V$ colors between $-0.2$ and 0, resulting in magnitude differences of 0.04--0.052.
Thus, we converted all $V_{\rm CSS}$ measurements to $V$ magnitudes by adding 0.046.

To identify flares from the nightly-binned CRTS light curve, we calculated the median of all the fluxes as the background value.
Then we selected $>5\sigma$ significant flux excess by requiring that the excess flux over the background exceeds five times the error.
We identified a flare when there are at least two consecutive data points with such $>5\sigma$ flux excess.
Based on this criterion, we detected two flares in the host galaxies of AT 2019azh and AT 2024pvu, occurring in 2005 and 2006, respectively (Figure~\ref{fig:LC}).
We referred to the CRTS flares as AT 2019azh.I and AT 2024pvu.I, and the corresponding ZTF flares as AT 2019azh.II and AT 2024pvu.II.
These two long-interval rTDE candidates were previously noticed by  \citet{Hinkle2021} and \citet{Langis2025}, although they did not provide substantial evidence that the CRTS flares are TDEs.

\subsection{Light curve analysis} \label{sec:2.2}

\begin{figure*}
\centering
  \includegraphics[width=\linewidth]{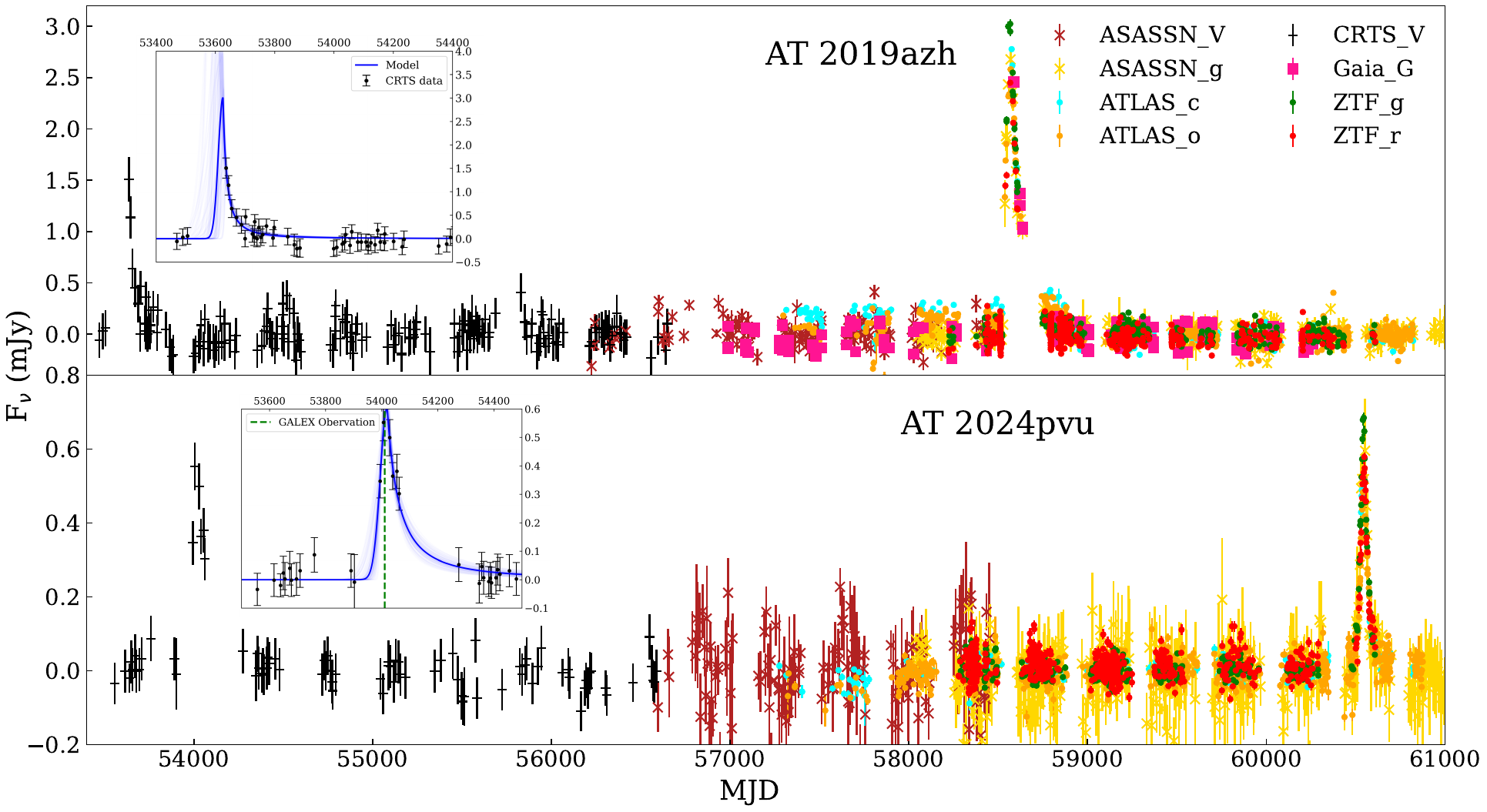}
  \caption{Host-subtracted optical light curves of AT 2019azh and AT 2024pvu.
  The insert panels show the partial view of the CRTS flares and the GP models that fit the data (MCMC).
  For AT 2024pvu, we labeled the observation time of GALEX using a green dashed line.}
\label{fig:LC}
\end{figure*}

\begin{table*}
\small
\centering
\caption{Parameters of the light curve models for the two rTDEs}
\begin{tabular}{ccccccccc}
\hline
\hline
IAU ID         & $t_{\rm peak}$     & lg$L_{V,\rm peak}$     & $t_{\rm 1/2,rise}$    & $t_{\rm 1/2,decline}\ (f)$ & $t_{\rm 1/2,decline}(p)$ & lg$T_{\rm bb}$ & lg$R_{\rm bb}$ & lg$L_{\rm bb}$ \\
               & (1)                & (2)                    & (3)                   & (4)                      & (5)                      & (6)            & (7)            & (8)            \\ \hline
AT 2019azh.I   & 53626$^{+12}_{-11}$& 43.27$^{+0.36}_{-0.29}$& 18.5$^{+20.3}_{-10.0}$ & 10.2$^{+5.1}_{-3.7}$     & 7.0$^{+3.5}_{-2.5}$      & -              & -              & -              \\
AT 2019azh.II  & 58561$\pm1$        & 43.14                  & 24.7$^{+1.3}_{-1.0} $ & 44.1$^{+1.1}_{-0.9}$     & -                        & 4.46           & 14.80          & 44.31          \\
AT 2024pvu.I   & 54019$\pm6$        & 43.25$\pm0.05$         & 27.8$^{+13.9}_{-8.8}$ & 33.6$^{+9.6}_{-7.8}$     & 23.5$^{+6.7}_{-5.6}$     & 4.29$\pm0.01$  & 15.00$\pm0.02$ & 44.03$\pm0.05$ \\
AT 2024pvu.II  & 60552$\pm1$        & 43.30$\pm0.01$         & 24.0$^{+0.4}_{-0.4} $ & 14.35$^{+0.3}_{-0.3}$     & -                        & 4.33$\pm0.01$  & 14.99$\pm0.01$ & 44.14$\pm0.02$ \\
Yao23.ZTF-TDEs & -                  & 42.33-44.55            & 6.4-51.8              & 5.2-86.5                 & -                        & 3.96-4.58      & 14.45-15.56    & 42.99-45.44    \\
\hline
\hline
\end{tabular}
\begin{tablenotes}
    \item (1) The peak time; (2) The peak $V$-band luminosity; (3) Rise time scales, expressed as $t_{\rm 1/2,rise}=1.177 \sigma_{\rm rise}$; (4),(5) Decline time scales assuming full TDE ($p=-5/3$) and partial TDE ($p=-9/4$) models, expressed as $t_{\rm 1/2,decline}=(2^{-1/p}-1) t_0$.
    (6)-(8) blackbody temperature, radius and luminosity at the peak time.
    For the two CRTS flares, we listed the best-fitting parameters of the GP model, while the blackbody parameters for AT 2024pvu.I were from SED fitting results using GALEX UV data.
    For AT 2024pvu.II, we listed the best-fitting parameters of the model of \citet{van2021}, and parameters for AT 2019azh.II and ZTF TDEs taken from \citetalias{Yao2023}.
    The lg$L_{V,\rm peak}$ of all ZTF TDEs are inferred from the peak bolometric luminosity and the temperature at the peak time.
\end{tablenotes}
\label{tab2}
\end{table*}

AT 2024pvu.I shows a light curve with a fast rise and a slow decline, conforming to those of TDEs, while only the declining phase of AT 2019azh.I was detected.
We fit the light curves using a model assuming a Gaussian rise and a power-law decline (hereafter GP model) expressed as:
\begin{equation}
    L(t)=
    \begin{cases}
    L_{\rm max} \times e^{-(t-t_{\rm peak})^2/2\sigma_{\rm rise}^2}& \text{ , $ t < t_{\rm peak} $ } \\
    L_{\rm max} \times (1+\frac{t-t_{\rm peak}}{t_0})^p& \text{ , $ t > t_{\rm peak} $ }
    \end{cases}
\label{eq:GP}
\end{equation}
where $t_{\rm peak}$ and $L_{\rm max}$ are the peak time and peak luminosity of the flare, $\sigma_{\rm rise}$ and $t_0$ imply the time scales of rise and decline, and $p$ is the powerlaw index of the decline.
We fit with the Markov Chain Monte Carlo (MCMC) code \texttt{emcee} \citep{emcee}.
We first attempted to set $p$ as a free parameter, but found that the data was too sparse to constrain it.
Thus, we adopted two reasonable $p$ values, $p=-5/3$ and $p=-9/4$ for the cases of full TDE \citep{Rees1988} and partial TDE \citep{Coughlin2019,Miles2020}, respectively, to constrain other parameters.
The model fits the data well, as shown in the inset panels of Figure~\ref{fig:LC}.
The inferred model parameters are listed in Table~\ref{tab2}.
We list the $t_{\rm 1/2,decline}$ assuming two different $p$ values, while the choice of $p$ hardly influences other parameters.
The peak $V$-band luminosity and rise and decline time scales are all within the range of ZTF TDEs of \citet[][hereafter \citetalias{Yao2023}]{Yao2023}.

We also investigated the light curve parameters of the corresponding ZTF flares for comparison.
We adopted the parameters for AT 2019azh.II from \citetalias{Yao2023}, and obtained the parameters for AT 2024pvu.II by fitting the ZTF, ATLAS and Swift UVOT \citep{Swift_UVOT2005} data (see details in the Appendix~\ref{sec:A 1.2}) with the model of \citet{van2021}.
The model assumes a bolometric light curve expressed as equation~\ref{eq:GP}, and a blackbody spectrum with $T_{\rm BB}$ linearly varying with time.
The parameters are also listed in Table~\ref{tab2}.
For both rTDE candidates, the peak luminosities of the two flares show no difference within the margin of error.
The time scales are also comparable, except that AT 2019azh.II has a longer decline time scale than AT 2019azh.I.
The time intervals are 13.2 and 17.1 years in the rest frame for AT 2019azh and AT 2024pvu, respectively.

To examine whether any additional flares occurred in these two galaxies between those discovered by CRTS and ZTF, we collected light-curve data from other surveys, including ASASSN, Gaia, and ATLAS.
We display all the light curves in Figure~\ref{fig:LC}, which show no additional significant flares.

\subsection{Multi-band analysis of AT 2024pvu.I} \label{sec:2.3}

AT 2024pvu.II was identified as a TDE due to its persistent blue color, the strong and broad He II emission line in the spectrum, and the lack of AGN-like host signatures \citep{Stein2024}.
\citet{Langis2025} first reported AT 2024pvu.I using CRTS data, but the literature lacks evidence to exclude the SN possibility and to verify its TDE nature.

Luminous UV emission and high blackbody temperature are essential features that distinguish TDEs from SNe \citep[e.g.,][]{van2020,Gezari2021}.
We noticed that the host galaxy of AT 2024pvu was fortunately observed by the Galaxy Evolution Explorer \citep[GALEX,][]{Martin2005} during the CRTS flare.
The observation time of MJD$=54010.5$ was coincidentally near the peak of the CRTS flare.
In both the images in the FUV and NUV bands, a point source is detected at the galaxy center with no extended feature, suggesting that the UV emission is more likely related to the flare than to the host galaxy.

\begin{figure*}
    \centering
    \includegraphics[width=0.49\textwidth]{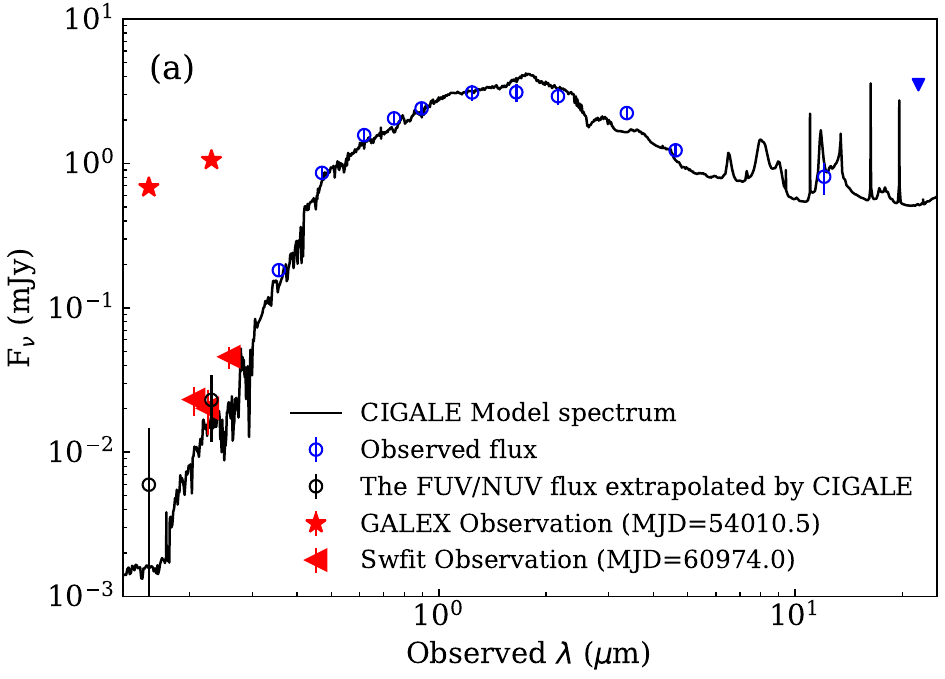}
    \includegraphics[width=0.49\textwidth]{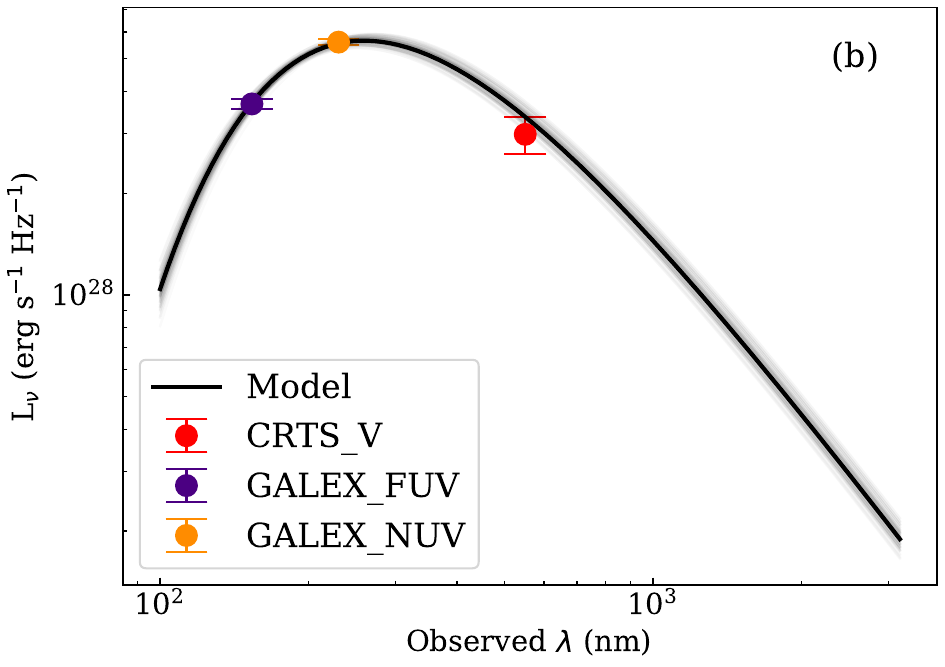}
    \caption{(a): The UV to MIR SED of AT 2024pvu.
    Blue points are the host galaxy's SED in the quiescent state, and black line is the best-fitting model from CIGALE with the minimum $\chi^2$.
    We show the GALEX and Swift UV photometries using red pentagrams and red triangles, respectively.
    The latter is consistent with the CIGALE's prediction (black open circle and the error bar), whereas the former shows a significant excess.
    (b): The SED of the CRTS flare at the time of GALEX observation, and the blackbody models that fit it (MCMC).}
\label{fig:AT2024pvu.I}
\end{figure*}

To verify whether the host galaxy is responsible for the UV emission, we triggered a new Swift ToO observation (ID:00016768015) in Nov 2025 (MJD=60974) after AT 2024pvu.II faded.
As can be seen from Figure~\ref{fig:AT2024pvu.I}(a), the NUV fluxes were several tens of times lower than that observed by GALEX.
To check for possible late-time UV emission of AT 2024pvu.II and measure the UV contribution of the host galaxy more accurately, we collected its archival multi-band photometric data, including Sloan Digital Sky Survey \citep[SDSS,][]{SDSS2000}, Two Micron All Sky Survey \citep[2MASS,][]{2MASS2006} and Wide-field Infrared Survey Explorer \citep[WISE,][]{WISE2010}, all were observed at times without flares (see Appendix~\ref{sec:A 1.2} for details).
We fit the spectral energy distribution (SED) using Code Investigating GALaxy Emission \citep[CIGALE,][]{CIGALE2019}.
In the fitting, we assumed that the star formation history can be described by the sum of two exponential decays and adopted the single stellar population templates of \citet{bc2003}.
We took dust attenuation/emission into account based on models of \citet{dustatt_CF2000} and \citet{dustemmision_dale2014}, while did not add any AGN component.
The resultant SED models of the host galaxy, if extrapolated to the UV bands, are consistent with the Swift UV observations in Nov 2025.
However, the synthetic fluxes in the FUV and NUV bands are about two orders of magnitude lower than observed values by GALEX, indicating that the transient source indeed dominates the UV emission at MJD$=$54010.5.

We obtained the SED of the flare at MJD=54010.5 using host-subtracted GALEX fluxes, and V-band flux inferred from the GP model that fits the CRTS light curve.
We fit the SED using the blackbody curve, and show the result in Figure~\ref{fig:AT2024pvu.I}(b).
We list the model parameters in Table~\ref{tab2}.
The blackbody temperature $T_{\rm BB}$ of $10^{4.29\pm0.01}$ K falls within the range of optical TDEs ($10^{3.96-4.58}$ K, \citetalias{Yao2023}), and is similar to that of the second flare, but significantly higher than that of SNe (typically $<10^4$ K).

We calculated the peak bolometric luminosity, the blackbody radius at the peak, and the total energy released assuming a constant temperature.
The peak luminosity of $L_{\rm bb}=10^{44.03\pm0.05}{\rm erg}$ ${\rm s^{-1}}$ is similar to that of the second flare.
The total energy released during AT 2024pvu.I can be estimated to be $E_{\rm UV,I} = 1.2 \times 10^{51} {\rm erg}$.

\section{Discussions}
\subsection{Nature of the CRTS flares: TDE or SNe?} \label{sec:3.1}

Could the two CRTS flares possibly be SNe?
As we have demonstrated, AT 2024pvu.I has a blackbody temperature that is too high to be interpreted as an SN.
In addition, its peak luminosity corresponds to an absolute magnitude of $M_V=-19.69\pm0.13$, higher than that of most SNe, which are generally fainter than $M_V=-19.4$ \citep[e.g.,][]{Li2011}.
Therefore, AT 2024pvu.I is unlikely to be an SN.
Besides, the peak absolute magnitude of AT 2019azh.I of $-19.70^{+0.72}_{-0.91}$ has too large an uncertainty to exclude an SN possibility.

We calculated the expected number of SNe that the CRTS observations could detect in the sample, and checked whether it accounts for the observed number of flares, which is two.
We describe in detail how we estimated the two essential parameters: the SN incidence rate in each galaxy based on its properties, and the effective monitoring duration (EMD) of CRTS observations.
The word ``effective'' here reflects the fact that the CRTS is not deep enough and its sampling is relatively sparse.

We first estimated the SN incidence rate.
The rate of SNe Ia is correlated with the stellar mass $M_\star$ and the star formation rate (SFR, $\dot{M_\star}$), and that of core-collapse (CC) SNe depends mainly on $\dot{M_\star}$.
Therefore, we estimated these two parameters by fitting the SEDs of the galaxies.
We collected SED data as described in detail in Appendix~\ref{sec:A 1.3}, and fit them with CIGALE using the same method as for AT 2024pvu in Section~\ref{sec:2.3}.
The resultant $M_\star$ and $\dot{M_\star}$ are listed in Table~\ref{tab1}.
The $M_\star$ are in the range of $10^{9.3-10.5}\ M_\odot$.
Only two galaxies in the sample, AT 2022bdw and AT 2018meh, show significant star formation activity, with $\dot{M_\star}$ of $\sim1.5$ and $\sim0.5\ M_\odot$ yr$^{-1}$, respectively.
They are also the only two galaxies that have $>3\sigma$ detection in the WISE 22-$\mu$m band.
The remaining are passive galaxies with specific star formation rate $\dot{M_\star}/M_\star\ \lesssim10^{-11}\ {\rm yr}^{-1}$.

We estimated the Ia SN rate (${\rm SNR_{Ia}}$) using the correlation of \citet{Smith2012}:
\begin{equation}
    {\rm SNR_{Ia}} = 1.05^{+0.16}_{-0.16} \times 10^{-10}\ 
    M_\star^{0.68^{+0.01}_{-0.01}}
    + 1.01^{+0.09}_{-0.09} \times 10^{-3}\ 
    \dot{M_\star}^{1.00^{+0.05}_{-0.05}}.
\end{equation}
The typical ${\rm SNR_{Ia}}$ in the sample is $\sim10^{-3}\ \rm galaxy^{-1}\ yr^{-1}$.
We also estimated the CCSN rate (${\rm SNR_{CC}}$) assuming that it is proportional to $\dot{M_\star}$ following \citet{Botticella2012}:
\begin{equation}
    {\rm SNR_{CC}}={\rm K_{CC}} \times \dot{M_\star},
\end{equation}
where ${\rm K_{CC}}$ is the proportionality constant, depending on the initial mass function (IMF).
We adopted ${\rm K_{CC}}=0.01$ by assuming the IMF of \citet{Salpeter1955}, calculated following \citet{Botticella2012}.
In the sample, the ${\rm SNR_{CC}}$ of AT 2022bdw and AT 2018meh is 0.015 and 0.005 $\rm galaxy^{-1}\ yr^{-1}$, respectively, and that of the remaining galaxies is $\lesssim10^{-3}\ \rm galaxy^{-1}\ yr^{-1}$.

We then estimated the EMD of each galaxy through simulations by answering the question: what is the range of peak time of an SN that CRTS could detect?
The EMD depends on the depth and cadence of CRTS observations, and is also affected by the peak luminosity and time scales of the SNe.
It would be longer if the SN has a higher peak luminosity or evolves more slowly.

\begin{figure*}
    \centering
    \includegraphics[width=0.49\textwidth]{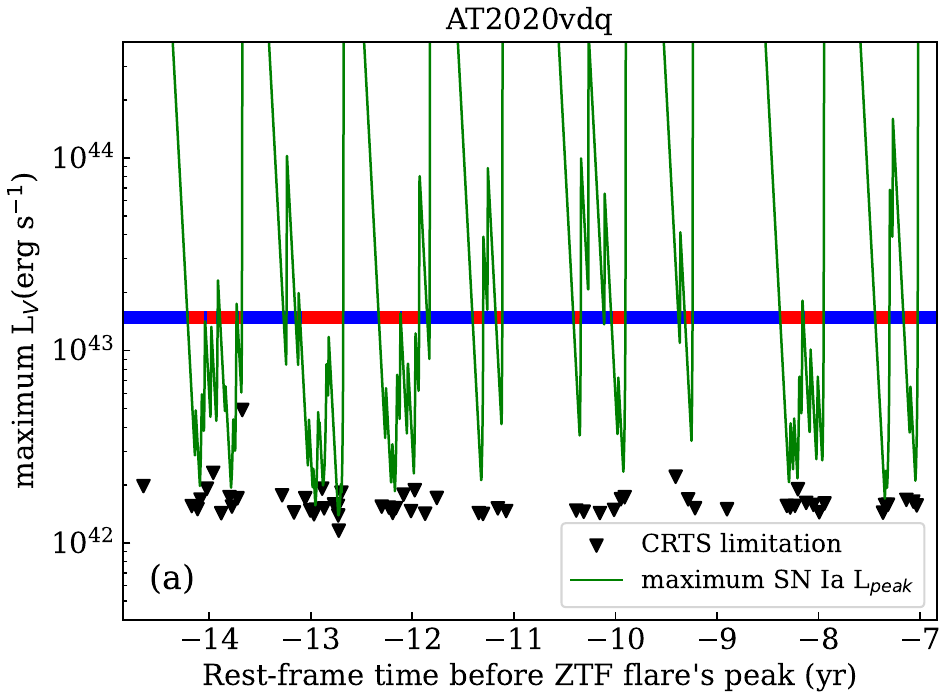}
    \includegraphics[width=0.49\textwidth]{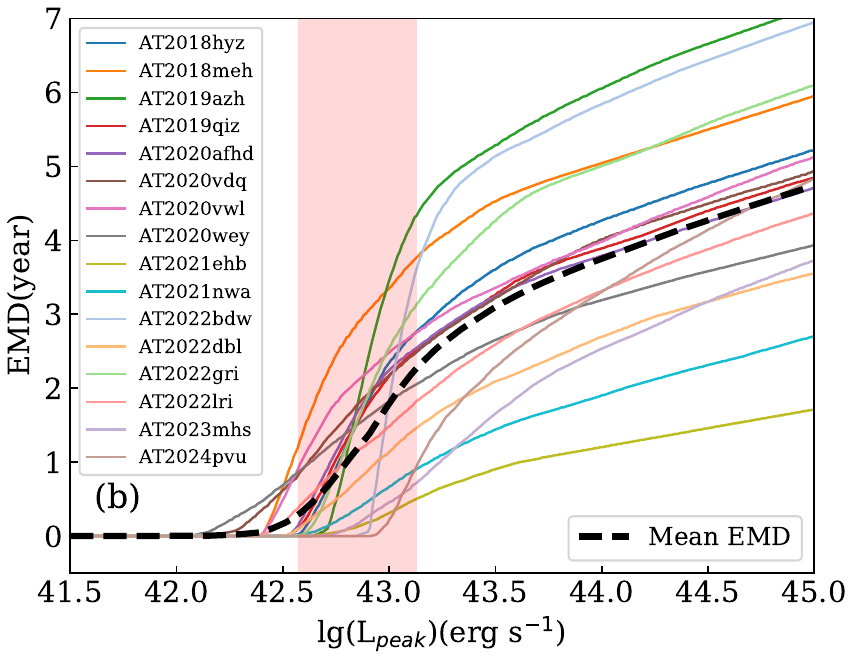}
    \caption{
    (a): An example using AT 2020vdq of how to calculate the EMD.
    We show the 5$\sigma$ upper-limit of each CRTS data point using black triangles, and the maximum $L_{\rm V}$ allowed by the observations as a function of $t_{\rm peak}$ using the green line.
    We converted the observational time to the phase before the ZTF TDE in the rest-frame.
    For a given $L_V$ (we show an example for $L_V\sim10^{43.14}$ erg/s, corresponding to $M_V\sim-19.4$), the time ranges when the maximum allowed $L_V$ is below this luminosity (red) are considered to have been effectively monitored, while others (blue) are not.
    (b): The resultant EMDs as functions of peak $L_V$ for the sample.
    We show the average EMD using the black dashed line, and the typical $L_{\rm peak}$ range of SN Ia ($-18<M_V<-19.4$) using the red shade.}
\label{fig:AT2020vdq_SNIa}
\end{figure*}

For SNe Ia, we assumed light curve models with a parabolic rise and an exponential decline, expressed as:
\begin{equation}
    L(t)=
    \begin{cases}
    L_{\rm peak} \times [1-(\frac{t-t_{\rm peak}}{t_{\rm rise}})^2]& \text{ , $ t < t_{\rm peak} $ } \\
    L_{\rm peak} \times e^{-(t-t_{\rm peak})/\tau}& \text{ , $ t > t_{\rm peak} $ }
    \end{cases}
\end{equation}
where $t_{\rm peak}$ and $L_{\rm peak}$ are the peak time and the peak $V$-band luminosity, and $t_{\rm rise}$ and $\tau$ represent the time scales of rise and decline.
We adopted $t_{\rm rise}=17\ {\rm days}$ and $\tau=16\ {\rm days}$ according to the statistical study of SN Ia \citep[e.g.,][]{Strovink2007,Desai2024}.
With these models with $t_{\rm peak}$ and $L_{\rm peak}$ as independent parameters, we generated simulated light curves of hypothetical SNe.
For each $t_{\rm peak}$ value, we checked above which $L_{\rm peak}$ value the hypothetical SN could be identified as a flare with two consecutive $>5\sigma$ excesses.
Figure~\ref{fig:AT2020vdq_SNIa}(a) shows how the minimum $L_{\rm peak}$ required for the SN detection (or the maximum $L_{\rm peak}$ allowed by the actual no detection) varies with $t_{\rm peak}$.
For a given $L_V$, the EMD is then calculated over the time range in which the maximum allowed $L_{\rm peak}$ is below this luminosity.
We present the inferred EMDs of the sample as a function of peak $L_V$ in Figure~\ref{fig:AT2020vdq_SNIa}(b).
The peak $M_V$ of SNe Ia is typically in the range from $-18$ to $-19.4$ \citep[e.g.,][]{Desai2024}.
We list EMDs for these two $M_V$ values in Table~\ref{tab1}, which are 0--1.2 and 0.5--4.3 years for $M_V$ of $-18$ and $-19.4$, respectively.

We finally estimated the expected number of SNe CRTS could detect in the sample.
By adopting EMD with $M_V=-19.4$ as an upper limit, we expected a total of $<0.068$ SNe Ia.
CCSNe are less luminous than SNe Ia with a typical $M_V$ of $>-18$, and we estimated the expected number of $\lesssim0.01$ for CCSNe using EMD for $M_V=-18$.
The estimation of CCSNe is highly uncertain due to the large dispersion of $L_{\rm peak}$ and the diverse shape of the light curve, which is generally different from SNe Ia.
Fortunately, it is unlikely that CCSNe dominate the SN detection in our sample for the following three reasons.
First, the typical $5\sigma$ depth of CRTS observations in our sample is $M_V=-17.9$, and only $\sim15\%$ of CCSNe at the bright end of the luminosity function are detectable with this depth \citep[e.g.,][]{Grayling2023}.
Second, only 2 out of 16 galaxies in our sample are star-forming galaxies with high CCSN rates, while the rest are passive galaxies.
Third, both CRTS flares are detected in passive galaxies, whereas none is detected in the two star-forming galaxies.

Assuming a Poisson distribution, the expectation of $\lesssim0.08$ leads to a probability of observing at least 2 flares of only 0.3\%.
We rule out the possibility that both flares are caused by SNe given the low probability.
Furthermore, the probability that one of the two flares is SN is only $\sim7\%$, so it is likely that neither of them is SN.

Could the two CRTS flares originate in AGN flares \citep[e.g.,][]{Graham2017}?
\citet{Hinkle2021} studied the pre-flare properties of AT 2019azh's host galaxy in detail and found that it was a post-starburst galaxy with no AGN features, such as broad emission lines, strong high-ionization narrow emission lines and X-ray emission.
We also found no evidence of AGN in the spectrum and archival X-ray observations of AT 2024pvu\footnote{We will present our analysis in detail in a subsequent work.}.
In addition, no AGN component is needed when fitting the SEDs with CIGALE for both galaxies.
The MIR color W1-W2 is $0.02\pm0.04$ and $0.08\pm0.04$ for AT 2019azh and AT 2024pvu, respectively, agreeing with inactive galaxies with W1-W2$<0.8$ \citep{Stern2012}.
Therefore, we exclude the possibility of AGN flares.

\subsection{Excessively high TDE rate} \label{sec:3.2}

In this subsection, we consider the implications for the TDE rate under the assumption that both CRTS flares are TDEs.
If assuming that the TDE rate in the galaxies in our sample is the average of $M_\star\sim10^{10}\ M_\odot$ galaxies, $3.2 \times 10^{-5}\ \rm galaxy^{-1}\ yr^{-1}$ \citepalias{Yao2023}, even without considering EMD and directly using the duration of CRTS observations, which is 7--8 years, the expected value is only $\sim4\times10^{-3}$, while 2 were actually detected.
Therefore, this sample must have a TDE rate well above the average.
We then quantitatively calculated the excess of the TDE rate relative to the average value.

\begin{figure*}
    \centering
    \includegraphics[width=0.49\textwidth]{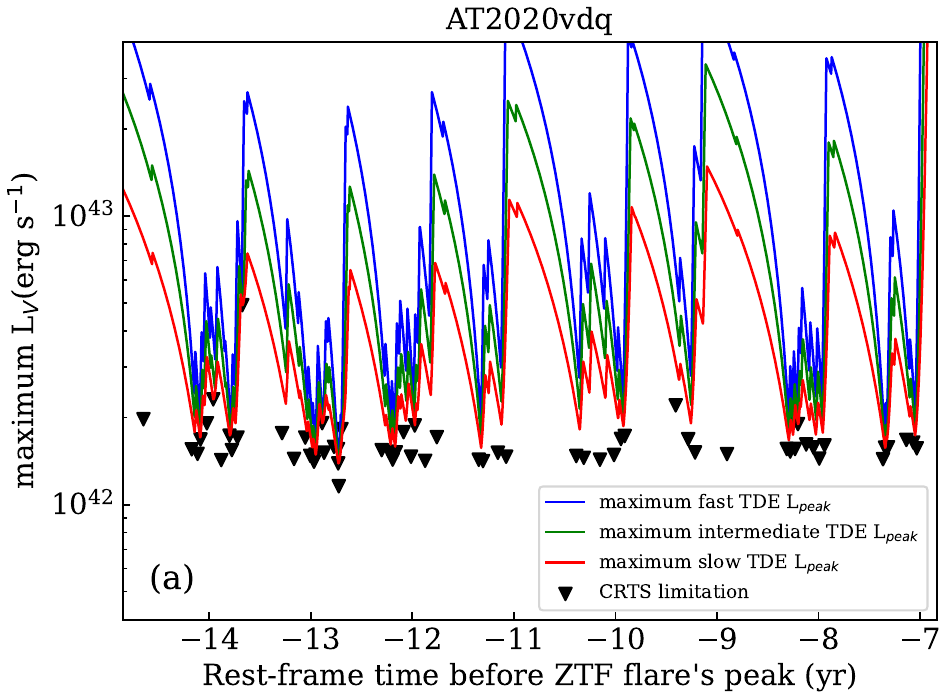}
    \includegraphics[width=0.49\textwidth]{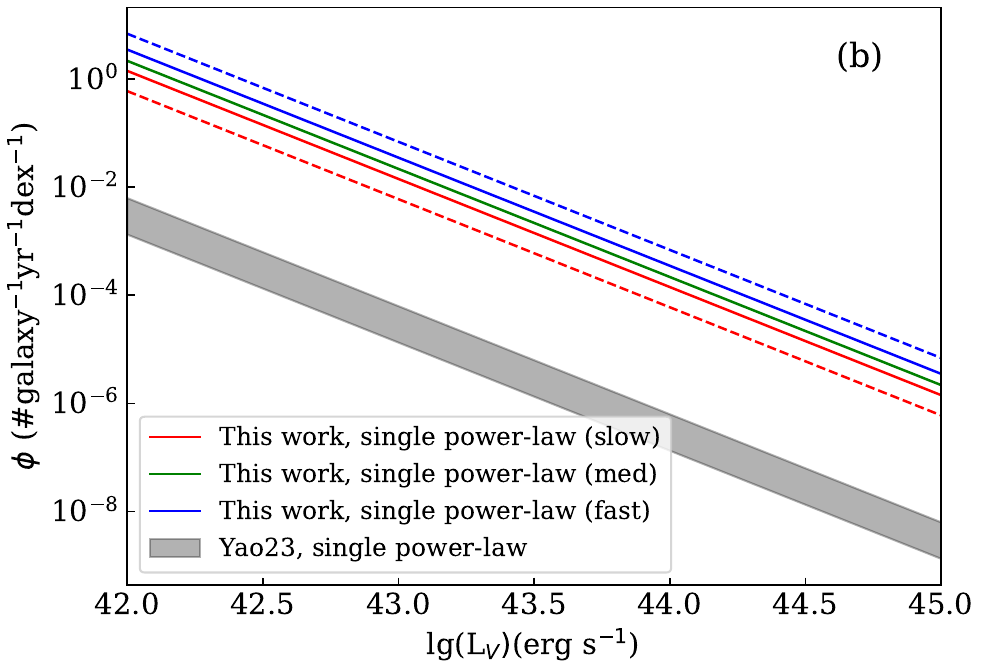}
    \caption{
    (a): The same example as Figure~\ref{fig:AT2020vdq_SNIa}(a), but for TDEs.
    (b): The TDE LF in our sample, with EMD calculated assuming different time scales (blue, green, and red for fast, intermediate and slow types, respectively).
    We present the best estimates in solid lines, and the $1\sigma$ upper limit for the fast type and the $1\sigma$ lower limit for the slow type, both shown in dashed lines.
    We also show the TDE LF from \citetalias{Yao2023} in grey shade for comparison.}
\label{fig:AT2020vdq_TDE}
\end{figure*}

We started by calculating EMDs for TDEs using the same method as used in Section~\ref{sec:3.1}.
We assumed GP light curve models expressed as equation~\ref{eq:GP}.
Due to the dispersion of the rise and decline time scales of ZTF TDEs \citepalias{Yao2023}, we considered three scenarios of $\sigma_{\rm rise}$ and $t_0$:
\begin{itemize}
\item[1.] Slow: $\sigma_{\rm rise}=20\ {\rm days}$, $t_0=100\ {\rm days}$; 
\item[2.] Intermediate: $\sigma_{\rm rise}=15\ {\rm days}, t_0=55\ {\rm days}$;
\item[3.] Fast: $\sigma_{\rm rise}=10\ {\rm days}, t_0=35\ {\rm days}.$
\end{itemize}
We show an example of how the maximum $L_{\rm peak}$ allowed by observations varies with $t_{\rm peak}$ in Figure~\ref{fig:AT2020vdq_TDE}(a), indicating that EMD increases as the TDE evolves more slowly.
We list the EMDs for $L_{\rm peak}=10^{43}$ erg s$^{-1}$ for each galaxy in Table~\ref{tab1}, which are in the range of 1--6 years for most galaxies.

For galaxies in our sample, we assumed a TDE luminosity function (LF) with a form of single power-law \citep{van2018,Yao2023}, expressed as:
\begin{equation}
    \phi({\rm lg}{L_V})=\dot{N_0} \times 10^{-\gamma({\rm lg}{L_V}-{\rm lg}{L_0})},
\end{equation}
where $L_V$ is the peak luminosity of TDE in the $V$-band, $\dot{N_0}$ is the normalization in unit of $\rm galaxy^{-1}\ yr^{-1}\ dex^{-1}$ at $L_0=10^{43}{\rm erg/s}$, and $\gamma$ is the powerlaw index.
Because we could hardly limit $\gamma$ with only two observed flares, we fixed $\gamma$ at 2 as \citetalias{Yao2023} obtained $\gamma=2.00^{+0.15}_{-0.14}$ for ZTF TDEs.
We constrained the only free parameter $\dot{N_0}$ with the two observed flares by using the Bayesian method as follows.

First, with the LF, the expected number of TDEs ($E_{\rm TDE}$) in an infinitesimal interval of $d{\rm lg}{L_V}$ can be calculated as:
\begin{equation}
    \frac{dE_{\rm TDE}}{d{\rm lg}{L_V}} = \phi({\rm lg}{L_V}) \times t_{\rm eff}({\rm lg}{L_V}),
\label{eq:E_TDE}
\end{equation}
where $\phi({\rm lg}{L_V})$ and $t_{\rm eff}({\rm lg}{L_V})$ are the LF and the total EMD of the whole sample at ${\rm lg}{L_V}$.
We divided the possible ${\rm lg}{L_V}$ range of 42--45 into bins of width 0.01 (we tried smaller steps and the result remained the same).
For the $i$-th bin, we calculated the expected number $E_i$ by integrating equation~\ref{eq:E_TDE} in this bin.
Meanwhile, the observed ${\rm lg}L_V$ distribution $\mathcal{N}$ can be expressed as:
\begin{equation}
    N_i=
    \begin{cases}
    1& \text{ , 43.25 or 43.27 is in the bin} \\
    0& \text{ , otherwise }
    \end{cases}
\end{equation}

Then, the probability $P(\mathcal{N})$ for producing the observed distribution $\mathcal{N}$ can be calculated using the Poisson distribution as:
\begin{equation}
    P(\mathcal{N}) = \prod_{i} P(N_i) = \prod_{i} \frac{E_i^{N_i}}{N_i!}e^{-E_i}
\end{equation}
Note that $P(\mathcal{N})$ is a function of the LF normalization $\dot{N_0}$.

Finally, assuming a logarithmically uniform prior distribution of $\dot{N_0}$, we calculated the posterior probability distribution of $\dot{N_0}$ using the Bayesian formula:
\begin{equation}
    \frac{dp(\dot{N_0}|\mathcal{N})}{d{\rm lg}\dot{N_0}} = \frac{P(\mathcal{N}|\dot{N_0})}{\int P(\mathcal{N}|\dot{N_0}) d{\rm lg}\dot{N_0}}
\end{equation}
The inferred $\dot{N_0}$ are $1.42^{+1.37}_{-0.82}$, $2.18^{+2.11}_{-1.26}$ and $3.52^{+3.40}_{-2.03}$ $\times 10^{-2}\ \rm galaxy^{-1}\ yr^{-1}\ dex^{-1}$ for slow, intermediate and fast types, respectively.
We show the resultant LFs in Figure~\ref{fig:AT2020vdq_TDE}(b).

To compare with the average TDE LF from ZTF, we converted the LF per volume reported by \citetalias{Yao2023} to that per galaxy.
This conversion requires the volume density of galaxies with $M_\star\sim10^{10}\ M_\odot$.
Using the galaxy's LF of \citet{Blanton2003} derived from $z\sim0.1$ SDSS galaxies, we obtained a volume density of $0.29-1.37 \times 10^{-2}\ {\rm Mpc^{-3}}$ by integrating with a lower bound of $M=-17$.
We show the converted TDE LF in Figure~\ref{fig:AT2020vdq_TDE}(b) in grey, which is 2-3 orders of magnitude lower than LF in our sample.
This indicates that for TDE host galaxies, the probability of another TDE occurring 5--19 years previously is much higher than the average.

\begin{table*}
\small
\centering
\caption{The properties of rTDEs in the sample}
\begin{tabular}{ccccccccccc}
\hline
\hline
ID         & $z$    & $\Delta t$ & lg$M_{\rm BH}$ & $E_1$ / $L_1$           & $E_2$ / $L_2$           & $r_{21}$ & $a$  & $\beta\ (1-e)$     & $t_{\rm next}$ & ref \\
           & (1)    & (2)        & (3)            & (4)                     & (5)                     & (6)      & (7)  & (8)                & (9)            & (10)\\ \hline
AT 2022dbl & 0.0284 & 1.9        & 6.4            & $E=1.6  \times 10^{50}$ & $E=7.9 \times 10^{49}$  & 0.5      & 209  & $3 \times 10^{-3}$ & 2026/01        & 1,2 \\
AT 2020vdq & 0.045  & 2.5        & 6.1            & $E=2.0  \times 10^{49}$ & $E=1.3 \times 10^{50}$  & 6.5      & 199  & $3 \times 10^{-3}$ & 2026/01        & 3   \\
AT 2019azh & 0.022  & 13.2       & 6.4            & $L_V=0.9-4\times10^{43}$& $L_V=1.4 \times 10^{43}$& $0.3-1.5$& 759  & $8 \times 10^{-4}$ & 2032           & 4   \\
AT 2024pvu & 0.048  & 17.1       & 7.9            & $E=1.2 \times 10^{51}$  & $E=6.6 \times 10^{50}$  & 0.55     & 2853 & $7 \times 10^{-4}$ & 2042           & 5   \\
\hline
\hline
\end{tabular}
\begin{tablenotes}
    \item (1) Redshift.
    (2) The time interval of two flares.
    (3) SMBH mass in unit of $M_\odot$.
    We took this data from the literature, except for AT 2024pvu.
    (4)(5) The total energy $E$ (or peak $V$-band luminosity $L_V$) of the first/second flare in unit of erg (erg s$^{-1}$).
    The total energy for AT 2019azh.I can hardly be constrained due to the lack of data, so we only provide $L_V$.
    (6) The ratio between the energy (or $L_V$) of the second and the first flares.
    (7)(8) The semi-major axis $a$ (in unit of AU) and eccentricity $e$ (given in $\beta\ (1-e)$) in the rpTDE model.
    (9) The peak time of the next flare predicted by the rpTDE model.
    (10) References are: 1. \citet{Lin2024}; 2. \citet{Makrygianni2025_2022dbl}; 3. \citet{Somalwar2025}; 4. \citet{Hinkle2021}; 5. \citet{Langis2025}.
\end{tablenotes}
\label{tab3}
\end{table*}

We estimated the fraction of rTDEs with intervals $\Delta t<20$ years in our sample.
Among the 16 ZTF TDEs, there are 4 rTDEs as listed in Table 3, including AT 2019azh and AT 2024pvu with $\Delta t$ of over ten years, and AT 2020vdq and AT 2022dbl with $\Delta t\sim2$ years.
Thus, the lower limit of the fraction is $\sim25\%$.
The average EMD of CRTS for TDEs with $L_V=10^{43}$ erg s$^{-1}$ is 3.2--4.3 years, covering 21\%--29\% of the time range of 5--20 years.
Assuming that the rTDE rate does not vary with $\Delta t$ and ignoring multiple repetitions in the same galaxy, we estimated that $\lesssim6$ rTDEs have been missed by observations.
Therefore, we concluded that $\sim25\%-60\%$ of optical TDEs are rTDEs.
We estimated the rate of rTDEs with $\Delta t<20$ yr following \citet{Somalwar2025} by scaling the TDE rate of \citetalias{Yao2023}:
\begin{equation}
    \begin{aligned}
    \mathcal{R}(rTDE) &\approx 25\%-60\% \times \frac{6\ \rm yr}{20\ \rm yr} \times 3.2 \times 10^{-5}{\rm\ galaxy^{-1}\ yr^{-1}} \\
     &\approx 2-6 \times 10^{-6}{\rm\ galaxy^{-1}\ yr^{-1}}
    \end{aligned}
\end{equation}
where 6 yr is the duration of the BTS TDE sample we used.
Both the fraction and the rate are higher than those of \citet{Somalwar2025} only considering $\Delta t<5$ years.

In our sample, there are at least two rTDEs with $5<\Delta t<20$ years, considering that more are likely to be missed, while there are only two with $\Delta t<5$ years.
Thus, most rTDEs may have relatively long time intervals.
Previous studies have overlooked this long-interval rTDE population, which may lead to statistical biases in rTDE research and an incomplete understanding of their origins.

\subsection{Origin of rTDEs}  \label{sec:3.3}

In this subsection, we discuss several possible origins of rTDEs and examine whether they can account for the observations of this sample.

\subsubsection{Two independent TDEs}

As we have demonstrated in Section~\ref{sec:3.2}, if all rTDEs in the sample are independent TDEs, the inferred TDE rate in the sample would be 2--3 orders of magnitude higher than the average.
Is this possible?

Post-starburst galaxies (or green valley galaxies) are widely known to have a higher TDE rate than the average.
AT 2019azh is hosted in a post-starburst galaxy \citep{Hinkle2021}.
In addition, the host galaxy of AT 2024pvu has a ${\rm NUV}-r$ color of $\sim4.5$, consistent with green valley galaxies \citep{Schawins2014,Belfiore2018}.
However, the TDE rate required to explain the rTDEs in these two galaxies is still $\sim1-2$ orders of magnitude higher than the average in post-starburst galaxies.

Theoretical studies have indicated that in some post-starburst galaxies, unequal-mass SMBHB about to merge can increase the TDE rate through the Lidov-Kozai mechanism or three-body scattering by up to several orders of magnitude \citep{Ivanov2005,Chen2009}.
It is generally thought that this mechanism can hardly be responsible for the overall high TDE rate in post-starburst galaxies, as this close SMBHB stage lasts only $\sim10^5$ years \citep[e.g,][]{Stone2020}.
However, the host galaxies of rTDE in our sample may be in this stage and have an extremely high TDE rate of $\sim10^{-2}\ \rm galaxy^{-1}\ yr^{-1}$ \citep{Liu2013_TDErate}.
The close SMBH companion might be detected using the gap in the TDE's X-ray light curve, which is caused by its gravitational disturbance to the TDE's debris fallback stream \citep{Liu2014,Shu2020}.
Although no such gap was observed in the X-ray monitoring of AT 2019azh \citep{Hinkle2021}, we cannot rule out its existence, as current X-ray monitoring is insufficiently covered.

The TDE rate decreases as the SMBH mass $M_{\rm BH}$ increases and is heavily suppressed at $M_{\rm BH}\gtrsim10^8\ M_\odot$, which is the Hills mass for Sun-like stars \citep[e.g.,][]{van2018,Polkas2024,Hannah2025}.
We adopted $M_{\rm BH}$ of $10^{6.44}$ for AT 2019azh from \citet{Hinkle2021}, and estimated $M_{\rm BH}\sim10^{7.9}\ M_{\odot}$ using the $M_{\rm bulge}-M_{\rm BH}$ correlation of \citet{Mc&Ma2013}.
For such a high $M_{\rm BH}$, even if there were a companion SMBH to increase the TDE rate, a considerable fraction of TDEs would not be observed as flares because the star falls into the event horizon before being tidally disrupted.
Thus, the high $M_{\rm BH}$ disfavors the scenario of independent TDEs for AT 2024pvu.

\subsubsection{Repeating partial TDEs}

We considered the scenario of repeating partial TDEs, which was proposed to explain periodic TDE-like flares, such as ASASSN-14ko, eRASSt J045650.3–203750, and AT 2023uqm.
In this model, the star moves in a highly eccentric orbit around SMBH with $R_p$ allowing for partial TDE ($\sim1-2\ R_t$ for fractional mass loss $\gtrsim10\%$, \citealt{Ryu2020_I,Ryu2020_III}), and triggers a flare each time it passes the periapsis.
Adopting the observed intervals in AT 2019azh and AT 2024pvu as the orbiting periods, we estimated $a$ to be 780 and 2850 AU using the $M_{\rm BH}$ and the Kepler's third law.
Assuming Sun-like stars, $R_t$ are 0.65 and 2.0 AU for AT 2019azh and AT 2024pvu, respectively.
Assuming $R_p=a(1-e) \approx 1-2\ R_t$, we estimated $1-e\approx10^{-3}$ for both rTDEs.

In the rpTDE model, the series of flares caused by the star should be regarded as a single event.
What impact will this have on the optical TDE event rate?

To answer this question, we need first to estimate the average number of flares ($\bar{N}$) produced by each rpTDE.
At present, we cannot measure $\bar{N}$ in our sample directly through observations.
We consider two possible ranges of $\bar{N}$: several, as in the case of eRASSt J045650.3–203750 (5), and several dozens, as in the case of ASASSN-14ko ($>30$).
The former requires a fractional mass loss of $\gtrsim10\%$, while the latter requires it to be $\lesssim5\%$.
We prefer the former for the following reasons.
Firstly, the flares' energies can vary several times in our sample (Table~\ref{tab3}), which does not match the situation in ASASSN-14ko, where the flares' energies are similar.
In the rpTDE model, the flares' energies may change significantly due to variations in the stellar structure.
For example, a decrease in stellar density after a pTDE may lead to an increase in the fractional mass loss of the next pTDE, or even lead to a full TDE \citep[e.g.,][]{Liu2025}.
Therefore, observing significant changes in the flare's energy may indicate that a series of rpTDEs is coming to an end, thereby supporting a lower number of repetitions.
Secondly, the average energy of the flares in our sample is on the order of $10^{51}$ erg (Table~\ref{tab3}), which is higher than that of ASASSN-14ko of $\sim10^{50}$ erg \citep{Huang2025} and comparable to the typical value of the ZTF TDE sample \citep{van2021,Yao2023}.
It would be more reasonable to assume $\bar{N}$ of several because $\bar{N}$ of several dozen requires disruptions of large mass stars with low probability.
Finally, for AT 2020vdq and AT 2022dbl, \citet{Somalwar2025} and \citet{Lin2024} found no earlier flares before the ZTF detection, implying that the ZTF detection is the first of the series of flares.
This situation would have too low a probability if assuming $\bar{N}$ of several dozen.
Therefore, it is reasonable to assume $\bar{N}$ of several for rpTDEs.

An rpTDE is more likely to fall within a time-limited sample than a single TDE.
This selection effect can lead to overestimation of the TDE rate and the rpTDE fraction.
There are a total of 18 ZTF flares in our sample considering that AT 2020vdq and AT 2022dbl each contributed 2 flares.
we assume that $\sim6-12$ of them were produced by rpTDEs, and the remaining $\sim6-12$ were produced by single TDEs.
By dividing the former by $\bar{N}$, which is assumed to be $\sim3-10$, we estimated the correction factor for the optical TDE rate to be $\sim40\%-80\%$ ($\sim2\times10^{-5}\ {\rm galaxy^{-1}\ yr^{-1}}$ using the observed value of the ZTF, \citetalias{Yao2023}), and the fraction of rpTDEs to be $\sim5\%-40\%$.
Taking into account rpTDEs with periods of $\gtrsim20$ years, the TDE rate would be further reduced, while the fraction of rpTDEs would be higher.

How do stars enter highly eccentric orbits with $1-e\approx10^{-3}$?
A widely accepted channel is that a close stellar binary system is tidally dissociated as it approaches the SMBH, with one star escaping and the other captured into a highly eccentric orbit \citep{Hills1988}.
The binary must be tight with separation $a_{\rm bin}\lesssim10\ R_\odot$ to survive stellar scattering in the galactic nucleus \citep[e.g,][]{Hills1988}.
The semi-major axis of the captured star is primarily related to $a_{\rm bin}$ \citep{Cufari2022}, which is roughly:
\begin{equation}
a \approx \frac{a_{\rm bin}}{2} \left( \frac{M_{\rm BH}}{M_\star} \right)^{2/3}.
\end{equation}
This correlation shows a scatter of 0.5--2 according to simulations \citep{YuAndLai2024}.
Assuming $M_\star=M_\odot$, the inferred $a_{\rm bin}$ is $\sim17$ and $\sim7$ $R_\odot$ for AT 2019azh and AT 2024pvu, respectively.
This is roughly in line with the condition of close binary, taking into account the scatter of the correlation, the measurement error of $M_{\rm BH}$, and the uncertainty of $M_\star$.

In the case of a close binary, the $R_p$ of the captured star is approximately equal to the $R_p$ of the original binary's center of mass.
A simple idea for interpreting rpTDE is that this initial $R_p$ allows partial disruption of the captured star ($R_p\lesssim2\ R_t$).
In this case, the first pTDE occurs when the binary system passes through the periapsis before dissociating.
Both stars may have experienced pTDE, but even so, only one flare could be observed because the time interval is short.
After that, every time the captured star returns to its periapsis, a flare occurs again until the star is fully disrupted or only an unbreakable core remains.

Can the fraction of rpTDEs of $\sim5\%-40\%$ be explained within this framework?
When estimating the proportions of different types of TDEs, we only consider single stars and close binaries because wide binaries in the galactic nucleus would transform into either single stars or close binaries after scattering with fellow stars.
We assumed that rpTDE can occur in close binaries with $R_p$ that allows partial TDE, whereas single TDE occurs in single stars or in close binaries with $R_p$ that allows full TDE.
According to the loss cone theory, the event rate of partial TDEs is roughly twice that of full TDEs \citep{Stone2020}, so the fraction of $R_p$ that meets the condition for partial TDE is about 2/3.
Thus, the observed rpTDE fraction further requires a fraction of close binaries in all stars $f_{\rm close}\sim7\%-60\%$.
This fraction is much higher than that observed in nearby open star clusters of only $\lesssim2\%$ ($a_{\rm bin}\lesssim10 R_\odot$, e.g., \citealt{Patience2002}).
Recent theoretical studies have shown that in galactic nuclei, gravitational perturbations from the SMBH and flyby stars can convert $\sim20\%-50\%$ of wide binaries into close binaries \citep{Dodici2025,HuangAndLu2025}.
Assuming an initial binary fraction of $\sim40\%-60\%$ \citep[e.g,][]{Duch2013}, these effects may increase $f_{\rm close}$ to $\sim10\%-30\%$, meeting the requirement to explain rpTDEs using the Hills mechanism.
This high fraction further predicts a large number of eclipsing binaries in the galactic nucleus, which can be tested observationally through infrared time-domain surveys.

Another way to boost the rpTDE rate in the Hills channel is to consider a larger $R_p$ of the binary's center of mass: the star is captured with $R_p>2R_t$ initially, with no partial TDE occurring; after orbiting for a period of time, the star's $R_p$ decreases or $R_t$ increases, and partial TDE can occur.
The $R_t$ of the captured star can increase through tidal heating and tidal spin-up \citep{Liu2025}.
In addition, the $R_p$ of the captured star can decrease under the gravitational perturbation of the nuclear cluster on a time scale of $\sim10^7-10^9$ yr, while $a$ remains almost unchanged \citep{PanAndLai2026}.
However, quantitative predictions of how much $R_t$ can increase or $R_p$ can decrease still require more detailed numerical simulations.

Besides the Hills mechanism, another channel to generate stars with highly eccentric orbits is the scattering of the S star cluster \citep{Bromley2012}, including internal scattering and scattering with intruding stars.
If AT 2019azh and AT 2024pvu are rpTDEs formed in this channel, the scattering position has a distance of $\sim10^3$ AU, coinciding with the inner edge of the S cluster in the Milky Way \citep{Eckart1996}.
However, the S cluster in the Milky Way is not dense enough to generate a large number of rpTDEs \citep{Alexander2017}.
Recently, \citet{JiangAndPan2025} proposed a model for interpreting QPE observed years after TDE occurs, involving collision between a pre-existing orbiting star and the newly formed TDE disk.
This model suggests exceptionally abundant stars in the vicinity of SMBH, which was explained as a result of recently faded AGN, where stars were captured by the AGN accretion disk \citep{PanAndYang2021} or formed in the disk in situ \citep{FanAndWu2023}.
Stars in such rich stellar disks may fall towards the SMBH in highly eccentric orbits through stellar scatters or the gravitational perturbation of a massive object, such as a companion SMBH.
The rpTDEs formed in this channel would have semi-major axes of $\sim10^3-10^4\ R_g$, the size of the AGN outer disk, consistent with the observations of our sample.
However, whether this mechanism can explain the observed rTDE rate remains to be studied in detail.

\subsubsection{Other possibilities and future observational tests}

Since only two flares have been detected for the rTDEs in our sample, we considered a phenomenon called ``double TDEs'' formed by sequential tidal disruptions of binary stars when they encounter the SMBH \citep[e.g,][]{Mandel2015,YuAndLai2024}.
However, for typical parameters, the time interval between two TDEs does not exceed several months, and the combined light curve would be similar to that of non-repetitive TDEs.
Double TDEs produced by SMBHB may have time intervals of up to several decades \citep[e.g,][]{Coughlin2018,Wu2018}.
However, the probability of a long interval ($>10$ years) is only $\lesssim1\%$, which cannot explain the high observed fraction of rTDEs.
Therefore, we excluded double TDEs as the origin of rTDEs in our sample.

In summary, we prefer to interpret rTDEs in our sample using the rpTDE model, while cannot rule out the possibility of independent TDEs.
The rpTDE model predicts that rTDEs are periodic, which requires verification through future observations.
In November 2025, a new flare was detected in IC 3599 \citep{IC3599_G2025}, which might belong to a series of TDEs with a period of $\sim18$ years, along with the two flares previously detected.
Follow-up observations of IC 3599 may provide conclusive evidence for the existence of rTDEs with a period of $>$10 years.
In addition, the independent TDE model predicts that the time intervals of rTDE are uniformly distributed.
This could not be tested using the current sample because of the large statistical uncertainty at the small sample size, and requires larger rTDE samples in the future.

\section{Summary}

Using CRTS data, we conducted a systematic search for rTDEs with time intervals of 5--19 years in a sample of 16 ZTF BTS TDEs with $z<0.05$.
Using a ``consecutive $>5\sigma$ excess'' threshold, we selected two CRTS flares in the host galaxies of AT 2019azh and AT 2024pvu.
The peak V-band luminosities ${\rm lg}L_{V,\rm peak}$ of the CRTS flares are $43.26^{+0.36}_{-0.29}$ and $43.25\pm0.05$, similar to those of the ZTF flares, and the time intervals are 13.2 and 17.1 years in the rest frame.

AT 2024pvu.I was fortunately observed by GALEX in the FUV and NUV bands near the peak.
The flare was detected in both bands, with UV fluxes two orders of magnitude higher than the host galaxy's contribution, inferred from Swift/UVOT data and optical-to-IR SED data.
The flare's UV/optical SED can be well described with a blackbody with a temperature of $\sim19500$ K, much higher than those of SNe and consistent with those of TDEs.

In addition, we estimated the expected number of SNe as follows.
We first predicted the rates of SNe Ia and CCSNe in the sample using galaxies' stellar masses and star formation rates.
We then calculated the EMDs for the CRTS observations assuming typical SN light curves, and found that the average EMD is 2.27 and 0.28 years for peak $M_V$ of $-19.4$ and $-18$.
We finally estimated the expected number of SNe Ia in all 16 galaxies to be $<0.068$, and that of CCSNe to be $\lesssim0.01$.
With a total expected number of $\lesssim0.08$, the probability that both flares in the sample are caused by SN is only $0.3\%$.
Therefore, it is likely that AT 2019azh and AT 2024pvu are rTDEs.

As the average EMD of the CRTS observations for TDEs is only 2.7--4.9 years, detecting 2 TDEs in 16 galaxies indicates that the TDE rate 5--19 years before the ZTF TDEs in this sample was extremely high.
Assuming a power-law LF, we estimated that the TDE rate is 2--3 orders of magnitude higher than the average.
Furthermore, we found that the fraction of rTDEs in the TDE sample is also high: adding another two rTDEs identified by ZTF with intervals of $\sim2$ years (AT 2020vdq and AT 2022dbl), there are at least 4 out of 16 rTDEs in this sample.
Considering possible rTDEs missed by the CRTS, the fraction of rTDEs with time intervals $<20$ years is estimated to be $\sim25\%-60\%$.

We proposed two interpretations for the high rTDE fraction.
One is that the two TDEs are independent, while the TDE rate in the galaxies in this sample was boosted by 2--3 orders of magnitude by a pc-scale SMBH companion.
The other is that both AT 2019azh and AT 2024pvu are rpTDEs of a star in a highly eccentric ($1-e\sim10^{-3}$) orbit.
In this case, the observed optical TDE rate is overestimated.
Assuming an average of several flares for each rpTDE, we estimated a correction factor for the rate of $\sim40\%-80\%$.
We prefer the rpTDE interpretation.
However, the channel through which the star enters such an orbit remains unclear.
It might be due to the Hills mechanism, or the scattering in an ultradense stellar disk left by a faded AGN.
We put forward predictions based on the two interpretations, which require future observational tests.

\begin{acknowledgements}
We thank the anonymous reviewer for helping to improve the manuscript.
We thank Zhen Pan for thoughtful suggestions on explaining the formations of rpTDEs.
This work is supported by the National Natural Science Foundation of China (NFSC, 12103002) and the University Annual Scientific Research Plan of Anhui Province (2022AH010013).
The CSS survey is funded by the National Aeronautics and Space Administration under Grant No. NNG05GF22G issued through the Science Mission Directorate Near-Earth Objects Observations Program.
The CRTS survey is supported by the U.S.~National Science Foundation under grants AST-0909182.
The ZTF are supported by the National Science Foundation under Grants No. AST-1440341 and AST-2034437 and a collaboration including current partners Caltech, IPAC, the Oskar Klein Center at Stockholm University, the University of Maryland, University of California, Berkeley, the University of Wisconsin at Milwaukee, University of Warwick, Ruhr University, Cornell University, Northwestern University and Drexel University. Operations are conducted by Caltech’s Optical Observatory (COO), Caltech/IPAC and the University of Washington at Seattle (UW).
We thank the Swift science operations team for accepting our ToO requests and arranging the observations.
This work made use of data supplied by the UK Swift Science Data Centre (UKSSDC) at the University of Leicester.

\end{acknowledgements}

\bibliographystyle{aa}
\bibliography{ref}

\begin{appendix}
\nolinenumbers

\section{Data collection and reduction}
\subsection{Milky Way galactic extinction} \label{sec:A 1.1}

For all the photometric data, we corrected for galactic extinction using python package \texttt{dust\_extinction} \citep{dust_ext2024}.
We adopted $\rm A_v$ values from the BTS catalog, and used models from \citet{F99} and \citet{G23} for UV bands and other bands, respectively.

\subsection{Light curve data} \label{sec:A 1.2}

Here we briefly introduce how we collected and reduced the light curve data other than CRTS, including ASASSN, Gaia, ATLAS, ZTF and Swift UVOT.

We obtained ASASSN light curves of AT 2019azh and AT 2024pvu in $V$ and $g$ bands from the website\footnote{\url{https://asas-sn.osu.edu/}} generated using image subtraction photometry.
Due to the larger flux error of ASASSN, we binned the data weekly.

We obtained the Gaia space telescope \citep[Gaia,][]{Hodgkin2021} $G$-band light curve of AT 2019azh from the alert system\footnote{\url{http://gsaweb.ast.cam.ac.uk/alerts/alert/Gaia19bvo/}}.
The Gaia alert system does not provide magnitude error.
However, AT 2019azh has $G\sim17$ and the typical magnitude error is $<0.05$ mag, ensuring the accuracy of the photometry.

We obtained the ATLAS light curves in $c$ and $o$ bands from the forced photometry server\footnote{\url{https://fallingstar-data.com/forcedphot/queue/}}.
We selected photometry on the difference images.
We removed data taken under bad weather (limit magnitude $<19$), and binned the light curve nightly.

We obtained the ZTF light curves in $g$ and $r$ bands from the NASA/IPAC infrared science archive\footnote{\url{https://irsa.ipac.caltech.edu/applications/Gator/}} (IRSA).
We removed bad data points using observational logs following the ZTF Science Data System explanatory supplement.
We subtracted the host flux using the median of the quiescent fluxes.

We downloaded the Swift UVOT data of AT 2024pvu from the NASA's Archive of Data on Energetic Phenomena\footnote{\url{https://heasarc.gsfc.nasa.gov/}}.
We made photometry on the UVOT images using \texttt{uvotsource} in High Energy Astrophysics Software.
We used an aperture with a radius of 5$\arcsec$, and calculated the background in two nearby circular regions without a source.

\subsection{SED data} \label{sec:A 1.3}

We collected the SED data of the whole galaxy as follows.

We obtained the GALEX photometries in the FUV and NUV bands from the Mikulski Archive for Space Telescopes\footnote{\url{https://galex.stsci.edu/GR6/}}.
We also downloaded the GALEX images of AT 2024pvu from the same archive.

We queried photometries from SDSS in the $u$, $g$, $r$, $i$ and $z$ bands using python package \texttt{astroquery} \citep{astroquery}, and selected model magnitudes.
For some galaxies that were not observed by SDSS, we collected the photometries from the Panoramic Survey Telescope and Rapid Response System \citep[PanSTARRS,][]{Panstar2016} in the $g$, $r$, $i$, $z$, and $y$ bands from the source catalog\footnote{\url{https://catalogs.mast.stsci.edu/panstarrs/}}, and selected the mean Kron aperture magnitudes.

We obtained the photometries of 2MASS in the $J$, $H$, and $K$ bands and WISE in the W1 to W4 bands through IRSA.
We adopted the 2MASS magnitudes in the extended source catalog if available; otherwise, we did not use 2MASS data in the SED fitting because the photometries in the point source catalog underestimate the whole galaxy's flux.
For the WISE magnitudes, we checked whether the pipeline identified the galaxy as a point source ($\texttt{ext\_flg}=0$).
If so, we adopted the magnitudes in the catalog from profile fitting.
Otherwise, i.e. the galaxy should be treated as an extended source, we measured Kron aperture fluxes on the WISE images using the python code \texttt{photutils} \citep{Photutils2025}, instead of using the magnitudes in the catalog.

\end{appendix}

\end{document}